\documentclass[%
 reprint,
 amsmath,amssymb,
 aps,
]{revtex4-1}

\usepackage{graphicx}
\usepackage{dcolumn}
\usepackage{bm}
\usepackage[T1]{fontenc}

\begin{document}

\title{How reactant polarization can be
used to change the effect of interference on reactive collisions}

\author{P. G. Jambrina}
 \email{pjambrina@usal.es}
\affiliation{%
 Departamento de Qu\'{\i}mica F\'{\i}sica ,
Universidad de Salamanca, Salamanca, Spain\
}%
\author{M. Men\'endez}
\author{A. Zanchet}
  \affiliation{
 Departamento de Qu\'{\i}mica
F\'{\i}sica I, Facultad de Ciencias Qu\'{\i}micas, Universidad Complutense de
Madrid , 28040 Madrid, Spain
}%
\author{E. Garc\'{\i}a}
 \affiliation{
 Departamento de Qu\'{\i}mica F\'{\i}sica, Universidad
del Pa\'{\i}s Vasco (UPV/EHU), Paseo de la Universidad 7, 01006 Vitoria, Spain
}%
\author{F. J. Aoiz}
 \email{aoiz@ucm.es}
 \affiliation{
 Departamento de Qu\'{\i}mica
F\'{\i}sica I, Facultad de Ciencias Qu\'{\i}micas, Universidad Complutense de
Madrid , 28040 Madrid  , Spain
}%

\date{\today}

\begin{abstract}
It is common knowledge that integral and differential cross sections (DCSs) are
strongly dependent on the spatial distribution of the molecular axis of the
reactants. Hence, by controlling the axis distribution, it is possible to
either promote or hinder the yield of products into specific final states or
scattering angles. This idea has been successfully implemented in experiments
by polarizing the internuclear axis before the reaction takes place, either by
manipulating the rotational angular distribution or by Stark effect in the
presence of an orienting field. When there is a dominant reaction mechanism,
characterized by a set of impact parameters and angles of attack, it is
expected that a preparation that helps the system to reach the transition state
associated with that mechanism will promote the reaction, whilst a different
preparation would generally impair the reaction. However, when two or more
competing mechanisms via interference contribute to the reaction into specific
scattering angles and final states, it is not evident which would be the effect
of changing the axis preparation. To address this problem, throughout this
article we have simulated the effect that different experimental preparations
have on the DCSs for the H + D$_2$ reaction at relatively high energies, for
which it has been shown that several competing mechanisms give rise to
interference that shapes the DCS. To this aim, we have extended the formulation
of the polarization dependent DCS to calculate polarization dependent
generalized deflection functions of ranks greater than zero. Our results show
that interference are very sensitive to changes in the internuclear axis
preparation, and that the shape of the DCS can be controlled exquisitely.
\end{abstract}

\maketitle

\section{Introduction}

One of the main goals of reaction dynamics is to provide tools that permits the control
of the outcome of a chemical event. \cite{Bernstein1987} Ideally, one would wish to set up an experiment where
only the desired product (in the desired internal state) will be produced. To achieve
that task, one should be able to imagine an experiment where the relative geometry and
energy of the incoming atoms can be selected in such a way that maximizes the
cross-section of the specific reaction whilst minimizing it for any other side-reaction.
To prepare such experiment, it would be necessary a complete knowledge of all the
competing dynamical mechanisms.

For reactions taking place in solution, the most it can be done is to choose
the temperature, the solvent and the catalyzers that would promote the
reaction. For bimolecular reactions in gas phase, however, it is possible to
carry out sophisticated experiments selecting, with extremely high resolution,
the relative energy, the initial quantum state, and to detect the products with
angular resolution. Moreover, it is possible to polarize the reactants
bond-axis or rotational angular momentum, so it can be achieved some control
into the relative geometry of the reactants before their interactions. For the
non-familiar reader that may sound utopia, but several groups have succeeded in
carrying out such experiments using either optical alignment
methods,\cite{Weida1997, Wang2011, Wang2012, Wang2016a, Wang2016b,Lorenz2001,
Brouard2015,Chadwick2014,Brouard2013,Kandel2000,Mukherjee2010,Mukherjee2011,PMZ:S17,PMZ:NC18,SLLMJACC:NC18}
brute force through intense laser fields
\cite{Loesch1992,Loesch1995,Loesch1998,Kim1996, Sakai1999,
Rosca2001,Stapelfeldt2003,Hamilton2005,Holmegaard2009,Ghafur2009,Friedrich1991a,Friedrich1999},
or using static orienting fields in tandem with hexapole state
selection.\cite{Parker1989,ABGNSW:PCCP15,DeLange2004,deLange1999,vanBeek2000a,VKKBOAGM:NC18,OGVAKNAGBM:NC17}

Based on the experimental set-up by Kandel {\em et al.},\cite{Kandel2000} in a
series of articles we have suggested a hypothetic crossed-beam experiment for
atom-diatom reactions where the diatom (in this case D$_2$) is prepared in a
$|v=0, j=2, m=0 \rangle$ state, where $v$, $j$, and $m$ are the vibrational,
rotational and magnetic quantum number.\cite{AMHKA:JPCA05} The latter is
determined with regard to a laboratory-fixed (space-fixed) quantization axis
(usually the light polarization vector or the direction of light propagation
for circularly polarized light). For a $m$=0 state, the internuclear axis is
preferably aligned along the quantization axis (a directed
state).\cite{KL:JPC87} The direction of the laboratory-fixed axis with regard
to the scattering frame -- defined by the initial relative velocity, $\bm k$,
and the plane containing $\bm k$ and $\bm k'$, the final relative velocity --
can be chosen arbitrarily, allowing us to select the relative geometry of the
incoming reactants.

Although the aforementioned experiments are indeed challenging, the idea behind
them it is rather simple.  The transition states have a well defined geometry;
therefore, if we select a configuration that aids the reactants to reach the
transition state, larger cross sections will be achieved, whilst if that
configuration does not lead to the neighborhoods of the transition state,
reaction will be impaired. However, that simple idea may not be valid when
quantum interference effects play an important role, and something apparently
as simple as a reaction mechanism is not so clearly defined. For example, for
the reactive collisions between H and D$_2$ at high collision energies
\cite{JHASJZ:NC15,JAASZ:CS16,SGZJMA:JCP16,AZ:PT18} and certain final
rovibrational states, the angular distribution showed prominent peaks and dips
that are the result of quantum interference between the classical mechanisms
previously described by Wrede and coworkers,\cite{GMW:JCP08,GMWA:JCP08} one
preferring collinear collisions with a linear intermediate collision complex
(denoted to as the {\em spiral} in ref.~\citenum{GMW:JCP08}), whilst the other
correlates with T-shape as intermediate triatomic arrangement (denoted in
ref.~\citenum{GMW:JCP08} as the {\em ear}).

If these mechanisms do not interfere, by selecting the geometry of the incoming reagents,
it would be possible to promote any of them. However, as long as they do, it does not
become easy to predict what would be the outcome of the collision upon selection of the
incoming geometry of the reactants. Throughout this article, we will tackle this problem
and we will show that an exquisite control of the position and intensity of the peaks of
the differential cross section (DCS) can be achieved by varying the polarization of the
reactants before the collision.

The article is organized as follows: In Section~\ref{Theory} we review the main
quantum mechanical (QM) expressions for the calculation of the observable DCS for a given
reactants preparation. Next, (Section~\ref{PDGDF}) we extend the formulation to the
treatment of Polarization Dependent Quantum Generalized Deflection Functions that will be
used to aid to the interpretation of the main features of the observable DCSs. The
results and their discussion are presented in Section~\ref{Res}. Finally, the main
conclusions of the article are highlighted in Section 4.

\section{Theory}\label{Theory}

\subsection{Polarization Dependent Differential Cross Sections}\label{PDDCS}

The polarization dependent differential cross sections (PDDCSs) are the polarization
moments that describe either the distribution of the initial angular momentum vector
(and, consequently, of molecular axes) that give rise to the reaction at a given
scattering angle into a final state, or the distribution of the rotational angular
momentum resulting from a reaction into a scattering angle. These magnitudes, their
expressions and physical meanings have been discussed at length in previous publications
\cite{SOZ:JPC95,ABE:JCP96,MABS:JCP99,AMHKA:JPCA05,AAM:PCCP08} and the reader is referred
to them for a thorough explanation. In the present context, we will consider the former;
that is, the PDDCSs associated with the ${\bm k}-{\bm j}-{\bm k'}$ correlation, also
known as reactant $j$-PDDCSs, which, for a given state-to-state process,  provide
information about the preferred direction of the rotational diatomic angular momentum
($\bm{j}$) of the incoming molecule to give rise to reaction into a specific final state.

The PDDCSs are {\em intrinsic} magnitudes, hence pure dynamical properties that reflect
the anisotropy of the potential and whose values will depend on the steric requirements
of the studied process. Therefore, intrinsic polarization moments are inherent to the
collision process and are independent of external circumstances (the experimental setup).
Accordingly, they can be directly extracted from the Scattering matrix.

The expression that relates the unnormalized PDDCSs with the elements of the
scattering matrix is:
\begin{eqnarray}\label{Ukq} \nonumber
 U^{(k)}_q (\theta) &=& \frac{1}{2 j + 1} \sum_{\Omega'}\sum_{\Omega_1 \Omega_2}  f_{v',j',\Omega' ~v, j,\Omega_1}(\theta) \times
\\ && f^*_{v',j',\Omega'~ v, j,\Omega_2} (\theta) \langle j, \Omega_1, k q | j
\Omega_2 \rangle
\end{eqnarray}
where $\theta$ is the scattering angle (that between $\bm k$ and $\bm k'$),
$U^{(k)}_q(\theta)$ denotes the unnormalized $j$-PDDCS of rank $k$ and
component $q$, $\langle j, \Omega_1, k q | j \Omega_2 \rangle $ is the
Clebsch-Gordan coefficient, and $f_{v',j',\Omega' ~v, j,\Omega}(\theta)$ is the
scattering amplitude for the state-to-state, $v, j,\Omega \to v', j',\Omega'$,
process. [In previous works,\cite{Kandel2000,AMHKA:JPCA05,AAM:PCCP08} we
have used the normalized PDDCS, $S^{(k)}_q(\theta)= (2\pi/\sigma_{\rm iso}) \,
U^{(k)}_q(\theta)$, where $\sigma_{\rm iso}$ is the integral cross section in
the absence of angular momentum polarization.]

The scattering amplitude can be written in terms of the Scattering matrix
elements, ${\textsf{S}}^J_{v',j',\Omega' v, j,\Omega}$, as
\begin{equation}\label{scattampl2}
f_{v',j',\Omega' v, j,\Omega} (\theta) = \frac{1}{2 i \, k_{\rm in}} \sum_J  \,
(2 J +1) d^J_{\Omega' \Omega}(\theta) {\textsf{S}}^J_{v',j',\Omega' v,
j,\Omega} \, ,
\end{equation}
where $k_{\rm in}$ is the reactant's wave-number and $d^J_{\Omega'\Omega}(\theta)$ is the
reduced Wigner-rotation matrix. In what follows, to simplify the notation, we will write
$f_{\Omega' \Omega}(\theta)$, implying definite values of initial $v$, $j$ and final
$v'$, $j'$ states.

Each $U^{(k)}_q(\theta)$, $k \in[0,2j]$ $q \in [-k,k]$, provides different physical and
``directional'' information about the polarization of the rotational angular
momentum.\cite{MABS:JCP99} Even $k$ moments are associated with alignment and odd $k$
moments with orientation. When $k$=0, the corresponding PDDCS, $U^{(0)}_0(\theta)$, is
nothing but the unpolarized (or isotropic) DCS.

\subsection{Observable Differential Cross Section}

The counterpart of the intrinsic polarization moments are the {\em extrinsic}
polarization moments, $A^{(k)}_q$, that provide information about the
preparation of the reactant rotational angular momentum; that is, about the
various possible experimental schemes for orienting or aligning reactant
molecules.\cite{AMHKA:JPCA05}

Whilst the intrinsic PDDCSs inform us about the steric preference of the reactions (what
the reaction ``wants''), the extrinsic polarization moments tell us about the rotational
angular momentum and molecular bond axis distributions in the asymptotic region, prior to
any interaction between the reactants (what we  experimentally ``offer'' to the
reaction).  As such, extrinsic polarization moments are a consequence of external
circumstances (the experimental setup) rather than the reaction itself.

Hence, it is evident that the DCS that can be  measured for a given experimental
preparation of the rotational angular momentum (hereinafter observable DCS) will be a
combination of both extrinsic and intrinsic polarizations. Of course, in the absence of
external reactant polarization, the measurement would yield the conventional, unpolarized
or isotropic DCS, which quantifies the two-vector $\bm{ k}-\bm{ k'}$ correlation. The
expression that relates the observable DCS to both types of moments
is:\cite{AMHKA:JPCA05,AJMRA:PCCP11}
\begin{eqnarray}\label{dcsalphabeta} \nonumber
\frac{d \sigma_{\alpha}^{\beta}}{d \omega} \equiv {\rm d}\sigma(\theta|
\beta,\alpha) &=& \sum_k \sum_{q=-k}^{k} (2 k + 1 )
\left[U^{(k)}_q(\theta)\right]^* \times \\ && A^{(k)}_0  C_{kq}(\beta,\alpha)
\end{eqnarray}
This expression contains the products of extrinsic and intrinsic polarization
moments of the same rank. Apart from these moments, eqn~\eqref{dcsalphabeta}
includes the modified spherical harmonics, $C_{kq}(\beta,\alpha)$, needed to
rotate the extrinsic moments $A^{(k)}_q$, defined in the laboratory frame, to
the scattering frame where the $U^{(k)}_q(\theta) $ are defined. The Euler
angles that connect both the laboratory and the scattering frame are $\beta$
(polar), and $\alpha$ (azimuthal). Therefore, the observable DCS depends on
those angles; {\em i. e.}, for a given extrinsic preparation, it depends on the
angles that define the laboratory with respect to the scattering frame. These
angles can be varied experimentally (for instance, changing the light
polarization vector with respect to the relative velocity) giving rise to
different preparations of the angular momentum.

As an example, a possible preparation for the D$_2$ molecule in $v$=0 (whose reaction
with H atoms will be considered in this work) could be to select $|j=2, m=0 \rangle$
state, which is a directed state.\cite{KL:JPC87} This can be achieved by pure rotational
Raman scattering by selecting the right pump and Stokes laser frequencies for stimulated
Raman scattering. By excitation via the $S(0)$ transition from D$_2$($v=0,j=0$), a
considerable excitation to D$_2$($v=0,j=2$) can be produced quite effectively by setting
the polarizations of the stimulated Raman pump and Stokes lasers parallel to each
other.\cite{Kandel2000,Mukherjee2010,Mukherjee2011}

In that case, the only non-vanishing polarization parameters, $A^{(k)}_q$, in the
laboratory frame are:
\begin{eqnarray} \nonumber
A^{(0)}_0 = 1, && \quad  A^{(2)}_0 = \langle 20, 20 | 20 \rangle =
 -\sqrt{\frac{2}{7}}, \\ &&  A^{(4)}_0 = \langle 20, 40 | 20 \rangle = \sqrt{\frac{2}{7}}
\end{eqnarray}
Varying the $\beta$ and $\alpha$ angles, the prepared state can be directed at
any arbitrary direction in the scattering frame.

\begin{figure*}
\centering
\includegraphics[width=0.90\linewidth]{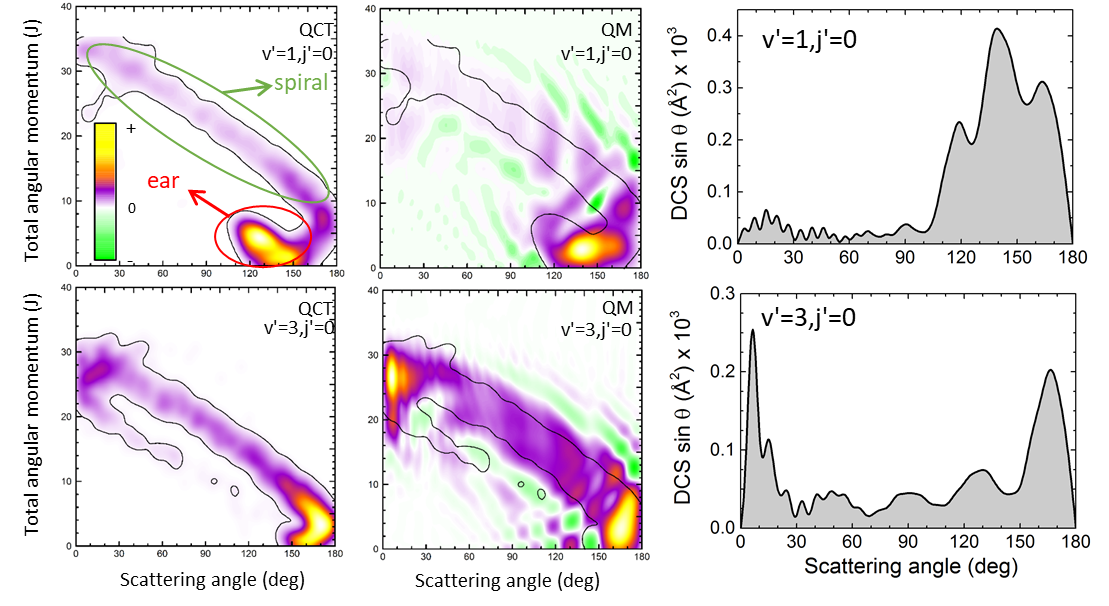}
\caption{Comparison of the QCT (left panels) and QM   (middle panels)
generalized deflection functions (GDF) for the  H +D$_2 (v=0,j=2)\rightarrow$
HD($v'$,$j'=0$) + D reaction ($v'$=1 upper panels, $v'$=3 bottom panels) at
$E_{\rm col}=$1.97 eV. The contour of the QCT deflection function has been
added to the QM $Q_r$ to make the comparison easier. The green contours
corresponds to destructive interferences ($Q_r < 0$). The QM DCSs for both
reaction channels are shown in the right panels.} \label{Fig1}
\end{figure*}

\subsection{Polarization Dependent Generalized Deflection Functions} \label{PDGDF}

The quantum deflection function is the functional of the deflection angle in terms of the
angular momentum $J$. It is a valuable magnitude to get some additional insight into a
scattering process, in particular to predict the presence of rainbows and interference
between near-side and far-side encounters.\cite{C:PCCP04,XC:JPCA09,SC:PCCP11}  As an
extension, we have recently derived a  Generalized Deflection Function
(GDF),\cite{JMA:CS18} that can be considered as the joint probability ({\em
quasi}-probability in the QM case) density function of $J$ and $\theta$. This
formulations allows us to plot a $J-\theta$ map that can be used to distinguish between
different reaction mechanisms, and to discern interference between them. The reader is
referred to Ref.~\citenum{JMA:CS18} for a detailed discussion.

In this article we extend the GDF formulation to deal with polarized reactants.
We start this section by summarizing the main expressions needed to calculate
QM GDFs that can be easily implemented in any QM scattering code. First of all,
we define the $J$-partial dependent scattering amplitude as:
\begin{equation}\label{scattj}
f^J_{\Omega' \Omega} (\theta) = \frac{1}{2i \, k_{\rm in}}   \, (2 J +1) d^J_{\Omega'
\Omega}(\theta) {\textsf S}^J_{\Omega' \Omega}
\end{equation}
where $\Omega$ and $\Omega'$ are bounded within $[-J,+J]$. The (summed over $J$)
scattering amplitude can now be written as:
\begin{equation}
f_{\Omega' \Omega} (\theta) =  \sum_{J=0}^{J_{\max}} f^J_{\Omega' \Omega}
(\theta) \,,
\end{equation}
where $J_{\max}$ is the maximum total angular momentum leading to reaction. The
DCS (that is, the $U^{(0)}_0(\theta)$ PDDCS) in the absence of reagent's
polarization ({\em i.e.}, isotropic preparation) can be expressed as a
function of the $J$-partial scattering amplitudes:
\begin{equation}\label{dcsscattj1}
{\rm d}\sigma_r(\theta|{\rm iso})=  \frac{1}{2j+1} \, \sum_{\Omega'}\sum_{\Omega}~
\,\sum_{J_1=0}^{J_{\max}} \sum_{J_2=0}^{J_{\max}}  f^{J_1}_{\Omega' \Omega}
(\theta)  f^{J_2 *}_{\Omega' \Omega} (\theta)\,,
\end{equation}
which can also be written as:\cite{JMA:CS18}
\begin{eqnarray}\label{dcsscattj2} \nonumber
{\rm d}\sigma_r(\theta|{\rm iso})&=&  \frac{1}{2j+1} \, \sum_{\Omega'}\sum_{\Omega}~
\sum_{J=0}^{J_{\max}}\,\sum_{J_1=0}^{J_{\max}} \sum_{J_2=0}^{J_{\max}} \frac{\left( \delta_{J_1,J}
+ \delta_{J_2,J}\right)}{2}\, \times \\ && f^{J_1}_{\Omega' \Omega}
(\theta)  f^{J_2 *}_{\Omega' \Omega} (\theta).
\end{eqnarray}

In a similar way, it is possible to express the PDDCSs as a function of the $f^J_{\Omega'
\Omega}$,
\begin{eqnarray}\label{Ukqj2} \nonumber
U^{(k)}_q(\theta)&=&  \frac{1}{2 j + 1} \sum_{\Omega'}~ \,
\sum_{J_1=0}^{J_{\max}} \sum_{J_2=0}^{J_{\max}} \sum_{\Omega_1} \sum_{\Omega_2}
f^{J_1}_{\Omega' \Omega_1} (\theta)  f^{J_2 *}_{\Omega' \Omega_2} (\theta) \times
\\ && \langle j \Omega_1, k q | j \Omega_2 \rangle \,,
\end{eqnarray}
By invoking the triangular relationship of the Clebsch-Gordan coefficients,
eqn.~\eqref{Ukqj2} can be simplified to:
\begin{eqnarray}\label{Ukqj3}
U^{(k)}_q(\theta) &=& \!\! \frac{1}{2 j + 1} \sum_{\Omega'}~
\sum_{J_1=0}^{J_{\max}} \sum_{J_2=0}^{J_{\max}} \sum_{\Omega}
f^{J_1}_{\Omega' \Omega}
(\theta)  f^{J_2 *}_{\Omega' \Omega + q} (\theta) \times \nonumber \\
&&\langle j \Omega, k q | j \Omega + q \rangle \,,
\end{eqnarray}
and, similarly to eqn~\eqref{dcsscattj2}, eqn~\eqref{Ukqj2} can be recast as
\begin{eqnarray}\label{skqj3}
U^{(k)}_q(\theta) &=&\frac{1}{2 j + 1} \sum_{\Omega'}~ \sum_{J=0}^{J_{\max}}
\sum_{J_1=0}^{J_{\max}} \sum_{J_2=0}^{J_{\max}} \sum_{\Omega} \frac{\left(
\delta_{J_1,J} + \delta_{J_2,J}\right)}{2} \, \times \nonumber \\ &&
f^{J_1}_{\Omega' \Omega} (\theta)  f^{J_2 *}_{\Omega' \Omega + q} (\theta)
\langle j \Omega, k q | j \Omega + q \rangle \,.
\end{eqnarray}

In a previous work,\cite{JMA:CS18} it was shown that it is possible to define a
quantum analog of the classical joint probability distribution of the
scattering angle and the total angular momentum, that can be written as:
\begin{eqnarray}\label{QJ} \nonumber
Q_r(\theta, J) &=& \frac{\sin \theta}{2 j + 1} \, \sum_{\Omega' \Omega} ~
\sum_{J_1=0}^{J_{\max}} \sum_{J_2=0}^{J_{\max}} \, \frac{\left( \delta_{J_1,J}
+ \delta_{J_2,J}\right)}{2} \times \\ && f^{J_1}_{\Omega' \Omega} (\theta)  f^{J_2
*}_{\Omega' \Omega} (\theta)\, .
\end{eqnarray}
Unlike the classical GDF and the classical or quantal DCSs, the QM GDF, $Q_r(\theta,J)$,
can take positive or negative values. Also, in contrast to the DCS, the QM GDFs are
additive with respect to the contribution of $J$ partial waves:
\begin{eqnarray}
\sum_{J=J_1}^{J_2} Q_r (\theta, J) + \sum_{J=J_2}^{J_3} Q_r (\theta, J) = \sum_{J=J_1}^{J_3} Q_r (\theta, J) \,,
\end{eqnarray}
with $J_1 < J_2 < J_3$.  Additionally, the QM GDF complies with the following
relationships:
\begin{eqnarray}
&\displaystyle{\sum_{J=0}^{J_{\max}} } Q_r (\theta, J)= {\rm d}\sigma_r(\theta |{\rm iso}) \sin \theta \\
&\displaystyle \int_{0}^{\pi} \int_{0}^{2\pi} Q_r (\theta, J) {\rm d}\theta {\rm d}\phi=
\frac{\pi}{k^2_{\rm in}}\,(2 J + 1) P_r(J)= \sigma^J_r\\
&\displaystyle \sum_{J=0}^{J_{\max}}~\int_{0}^{\pi} Q_r (\theta, J) {\rm d}\theta {\rm d}\phi= \sigma_{\rm iso}
\end{eqnarray}
where $P_r(J)$ is the reaction probability as a function of the total angular momentum.
It is related to the $J$-partial cross section by $\sigma^J_r=\pi/k^2_{\rm in}\, (2J+1)
P_r(J)$.

Following the same procedure as that used to derive the QM GDF, higher-rank
GDFs can be defined as:
\begin{eqnarray}\label{QJkq}
Q^{(k)}_q(\theta,J) =& \displaystyle{\frac{\sin \theta}{2 j + 1} }\, \sum_{\Omega'} ~
\sum_{J_1=0}^{J_{\max}} \sum_{J_2=0}^{J_{\max}} \, \sum_{\Omega} \, \frac{\left( \delta_{J_1,J}
+ \delta_{J_2,J}\right)}{2} \, \, \nonumber\\  & f^{J_1}_{\Omega' \Omega} (\theta)
f^{J_2*}_{\Omega' \Omega+q}(\theta)  \, \langle j \Omega, k q | j \Omega + q \rangle \, .
\end{eqnarray}
where $Q^{(k)}_q(\theta,J)$ denotes the GDF with rank $k$ and component $q$.

For $q$=0 moments, the only differences between eqn~\eqref{QJkq} and \eqref{QJ} is the
presence of the Clebsch-Gordan coefficient. In particular, for $k$=0 eqn~\eqref{QJ} is
recovered. Furthermore, integration of $Q^{(k)}_{q=0}(\theta,J)$ over $\theta$ allows us
to recover the $J$-dependent polarization parameters.\cite{AHJAJZ:JPCL12} For $q \neq 0$,
eqn~\eqref{QJkq} involves the product of $J$-partial dependent scattering amplitude of
different $\Omega$ values and no simple analytical formula can be derived, although
numerical integration leads to the $J$-dependent polarization parameters, $P^{(k)}_q(J)$.
Finally, regardless of the value of $q$, summation of eqn~\eqref{QJkq} over $J$ retrieves
the $U^{(k)}_q(\theta)\, \sin \theta$.

It is a common practice in stereodynamics to define renormalized PDDCSs, independent of
the flux at each scattering angle, and whose limits are well defined.\cite{AAM:PCCP08} In
principle, the same normalization could be applied to the the $Q^{(k)}_{q}(\theta,J)$.
However, because GDFs can take  positive or negative values (that could even canceled out
for a certain range of $J$s), renormalized $Q^{(k)}_{q}(\theta,J)$ have no definitive
limits and its use is not practical.

In analogy with the definition of the observable DCS, eqn~\eqref{dcsalphabeta}, the
$Q^{(k)}_{q}(\theta,J)$ can be combined to calculate $Q^{\beta}_{\alpha}(\theta,J)$, {\em
i.e.},  QM GDFs that depend on the experimental preparation of the reactants and whose
sum over $J$ yields the observable DCS. The expression for the calculation of
$Q^{\beta}_{\alpha}(\theta,J)$ is
\begin{equation}\label{Qalphabeta}
Q^{\beta}_{\alpha}(\theta,J) = \frac{\sigma_{\rm iso}}{2 \pi}
\sum_k \sum_{q=-k}^{k} (2 k + 1 ) \left[Q^{(k)}_{q}(\theta,J)\right]^* A^{(k)}_q  C_{kq}(\beta,\alpha)
\end{equation}
We will make extensive use of this equation to determine the QM GDFs under different
preparation of the angular momentum direction.

\section{Results and Discussion}\label{Res}

\label{Results}

\begin{figure}
\centering
  \includegraphics[width=0.9\linewidth]{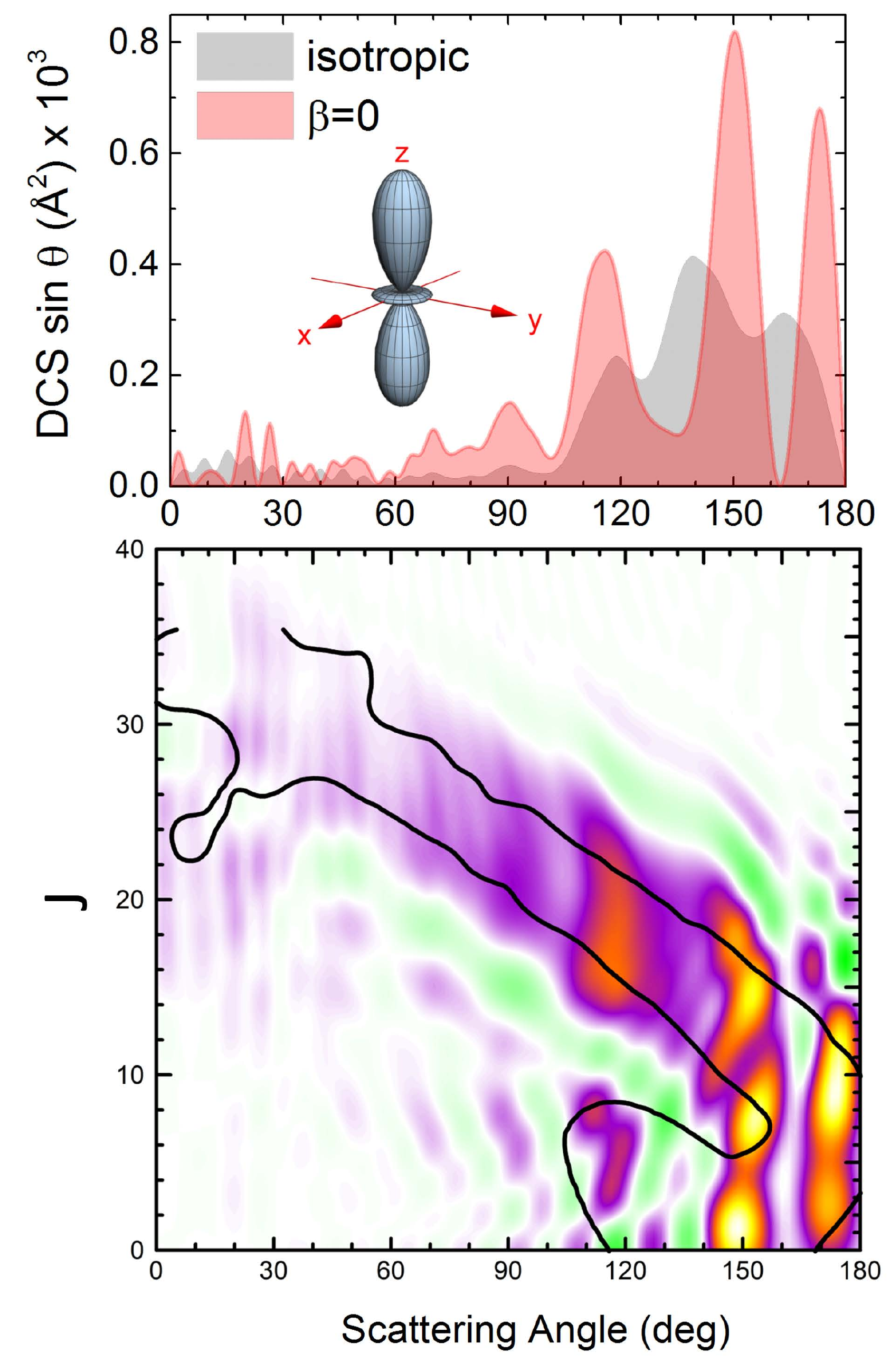}
\caption{Differential Cross Sections for the
{H+D$_2($v$=0,$j$=2\rightarrow$HD($v'$=1,$j'=0$)+D} reaction at $E_{\rm
col}=$1.97 eV for $\beta$=0. The isotropic DCS is shown in gray and the
observable DCS in red. The extrinsic preparation of the internuclear axis of
D$_2$ referred to the scattering frame is shown in an inset. The QM GDFs for the polarized reactants is shown in the bottom panel. }
\label{Fig2}
\end{figure}

We will start this section with the discussion of the shape of the isotropic QM
DCSs and its connection with the quasiclassical deflection functions (QCT GDF);
that is, the joint probability density function of $\theta$ and $J$. These
results have been presented previously\cite{JHASJZ:NC15,JAASZ:CS16} and are
repeated here for completeness and to link up with the subsequent discussion.
The interested reader is referred to Ref.~\citenum{JHASJZ:NC15,JAASZ:CS16} for
a thorough discussion.

The DCSs for the  H + D$_2$($v$=0,$j$=2) $\rightarrow$ D + HD($v'$,$j'$)
reaction at E$_{\rm col}$=1.97 eV are shown in the right panels of
Fig.~\ref{Fig1}. For $v'$=1,~$j'$=0, (upper panel) the products are mainly
scattered in the backward hemisphere where three peaks are observed, one peak
at 150$^{\circ}$, and two smoother peaks at 120$^{\circ}$ and 175$^{\circ}$.
These peaks were also present in the DCSs of the reactions of H with
D$_2$($v$=0,$j$=0) and with D$_2$($v$=0,$j$=1).\cite{JAASZ:CS16} The origin of
these structures was attributed to quantum interference between (essentially)
two competing classical mechanisms that can be easily elucidated with the help
of the QCT DF, which is shown in the left upper panel of Fig.~\ref{Fig1}. As
can be seen, in the backward region, $\theta > $90$^{\circ}$, there are two
separated regions that give rise to products at the same scattering angles. On
the one hand, there is scattering from the wide band running diagonally across
the $\theta$-$J$ map with a negative slope, that includes all $J$ values
contributing to the reaction but the smallest ones ($J \in [0,8]$), region
called {\em spiral}.\cite{GMW:JCP08} Although this region covers the whole
range of scattering angles, at backward angles the intensity is somewhat
larger. On the other hand, scattering in the backward region is mainly caused
by another, well separated, confined region, including only small $J$ values
(small impact parameters), $J \le $8, giving rise to products in a limited
range, 110-170$^{\circ}$, of scattering angles, region called {\em
ear}.\cite{GMW:JCP08} These two regions in the $\theta$-$J$ map are
characterized by distinct classical mechanisms (different impact parameters,
angle of attack, intermediate collision complex, etc.) that nevertheless give
rise to scattering at the same angles and into the same final states.
Therefore, there is small wonder that interference from these separate
contributions gives rise to oscillations in the DCS that explain the peak
structure observed.

Although the QCT GDFs for different initial $j$ states are remarkably
similar,\cite{JAASZ:CS16} the sharpness of the backward peaks in the DCS was
found to decrease with increasing $j$. The reason is not the progressive
quenching of interferences with increasing $j$, but rather the incoherent
superposition of contributions from different helicity states of the reagents
(for $j$=2, $\Omega= 0,\pm 1,\pm 2$). As a matter of fact, the decomposition of
the QCT GDF into the various $\Omega$ values shows that their respective
contributions are associated with different dynamical
mechanisms.\cite{JAASZ:CS16}

The QM GDF, depicted in the middle, upper panel of Fig.~\ref{Fig1}, has the advantage of
allowing the observation of interference between different mechanisms. As can be seen,
the {\em spiral} mechanism, associated with the diagonal band, is chopped into three
pieces, each corresponding to each of three peaks of the DCS. It is worth mentioning
that, in a clear contrast to what was observed for the reaction with D$_2$ in
$j$=0,\cite{JMA:CS18} there are no relevant destructive interferences (negative values of
$Q_r(\theta, J)$, represented as green stripes) between the various groups of $J$ . The
lack of negative values precludes the appearance of deep minima as those observed in the
DCS for $j$=0 (see Fig.~2 of Ref.~\citenum{JAASZ:CS16}) and makes the whole structure
much smoother.

\begin{figure}
\centering
  \includegraphics[width=0.8\linewidth]{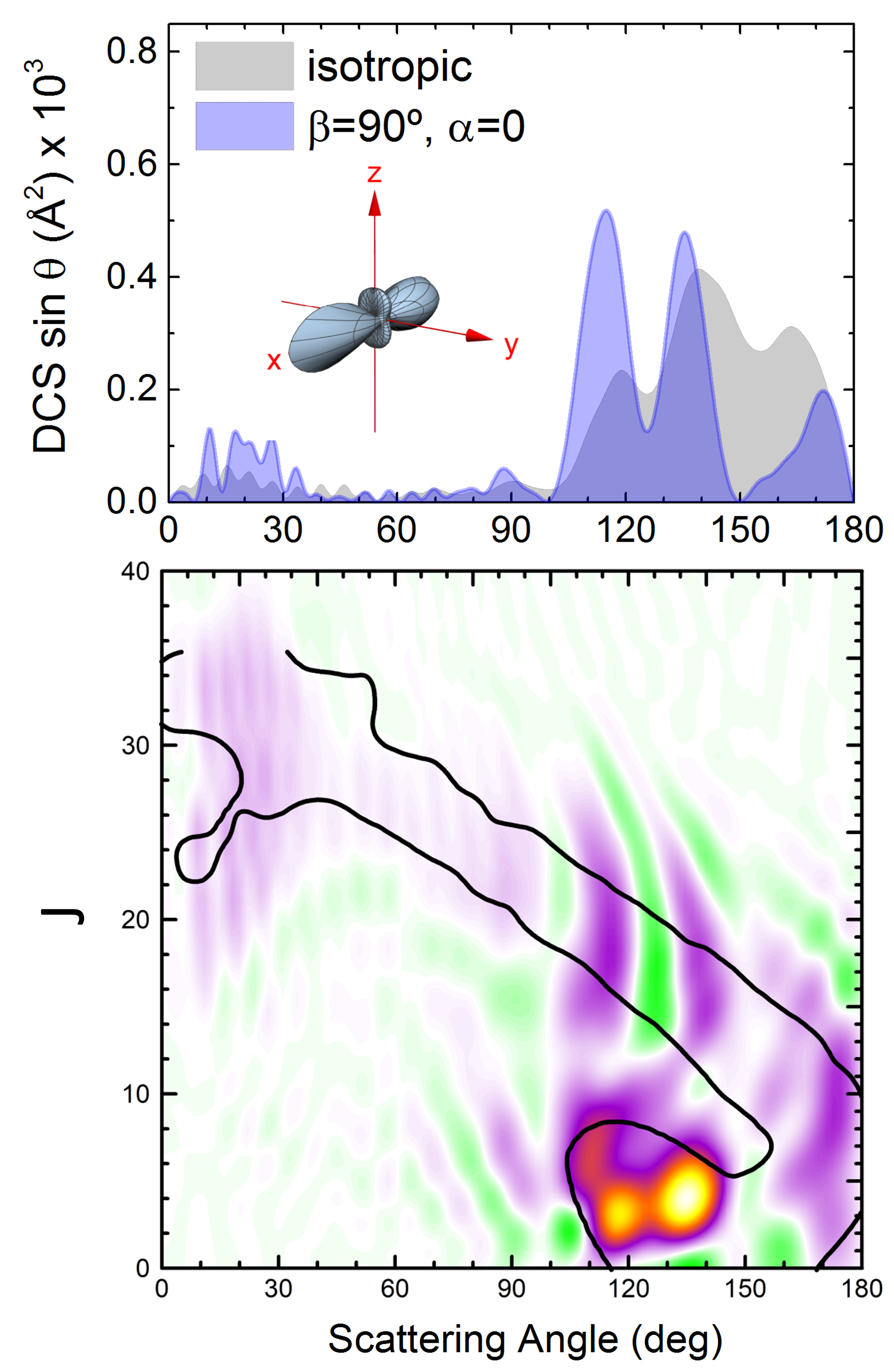}
\caption{ As Fig. \ref{Fig2} but for a $\beta=90^{\circ}$, $\alpha=0$
preparation.  } \label{Fig3}
\end{figure}

Fig.~\ref{Fig1} also contains the respective results for the H +
D$_2$($v$=0,$j$=2) $\rightarrow$ D + HD($v'$=3,$j'$=0), shown in the lower
panels. Although the DCS exhibits several oscillations in the whole range of
scattering angles, the observation of the classical and quantal GDFs rules out
the presence of strong interference as that observed for $v'$=1. Both QCT and
QM GDFs are very similar and can be ascribed to the pattern expected for direct
reaction with essentially a single mechanism: that corresponding to the {\em
spiral}, with large (small) angular momenta --or impact parameters-- associated
with forward (backward) scattering. Moreover, except for some details, the QM
GDF for the reaction with D$_2$ in $j$=0 and in $j$=2 are very much alike.

\begin{figure}[t!]
\centering
  \includegraphics[width=0.8\linewidth]{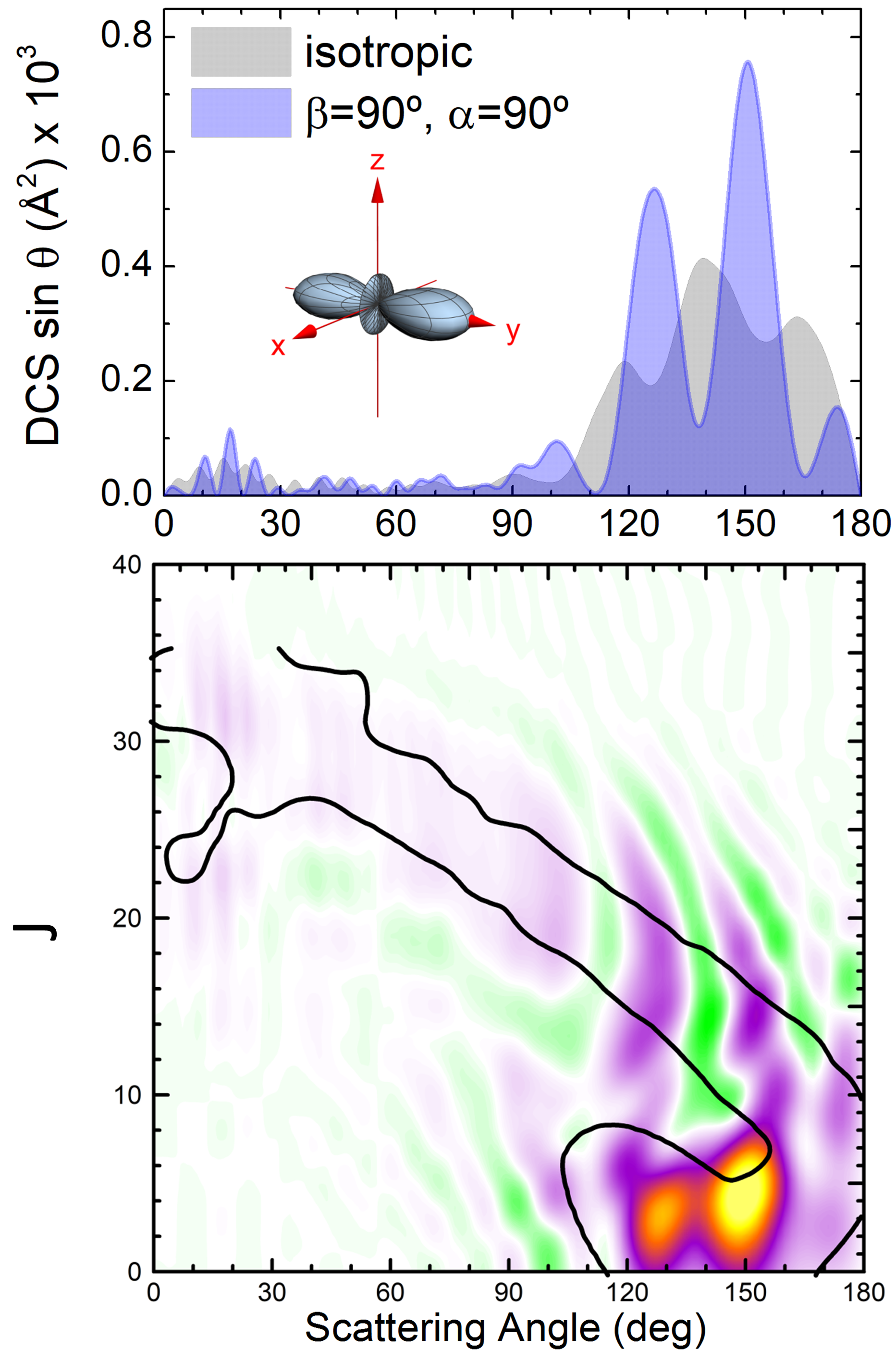}
\caption{As Fig.~\ref{Fig2} but for a $\beta=90^{\circ}$, $\alpha=90 ^{\circ}$
preparation.} \label{Fig4}
\end{figure}
\begin{figure}[hb!]
\centering
  \includegraphics[width=0.8\linewidth]{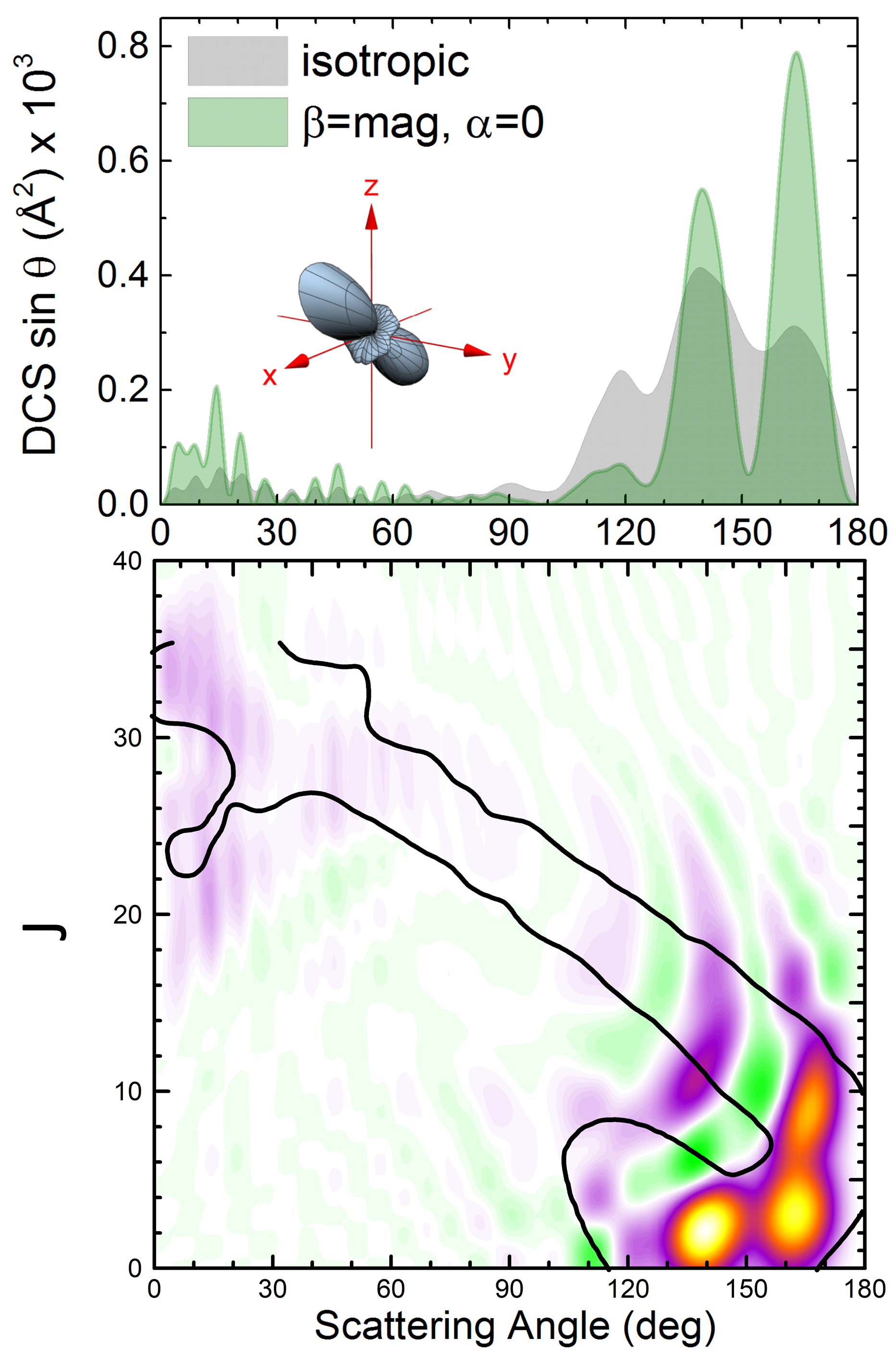}
\caption{As Fig. \ref{Fig2} but for a $\beta={\rm mag}$, $\alpha=0 $
preparation.} \label{Fig5}
\end{figure}
%
\begin{figure}
\centering
  \includegraphics[width=0.8\linewidth]{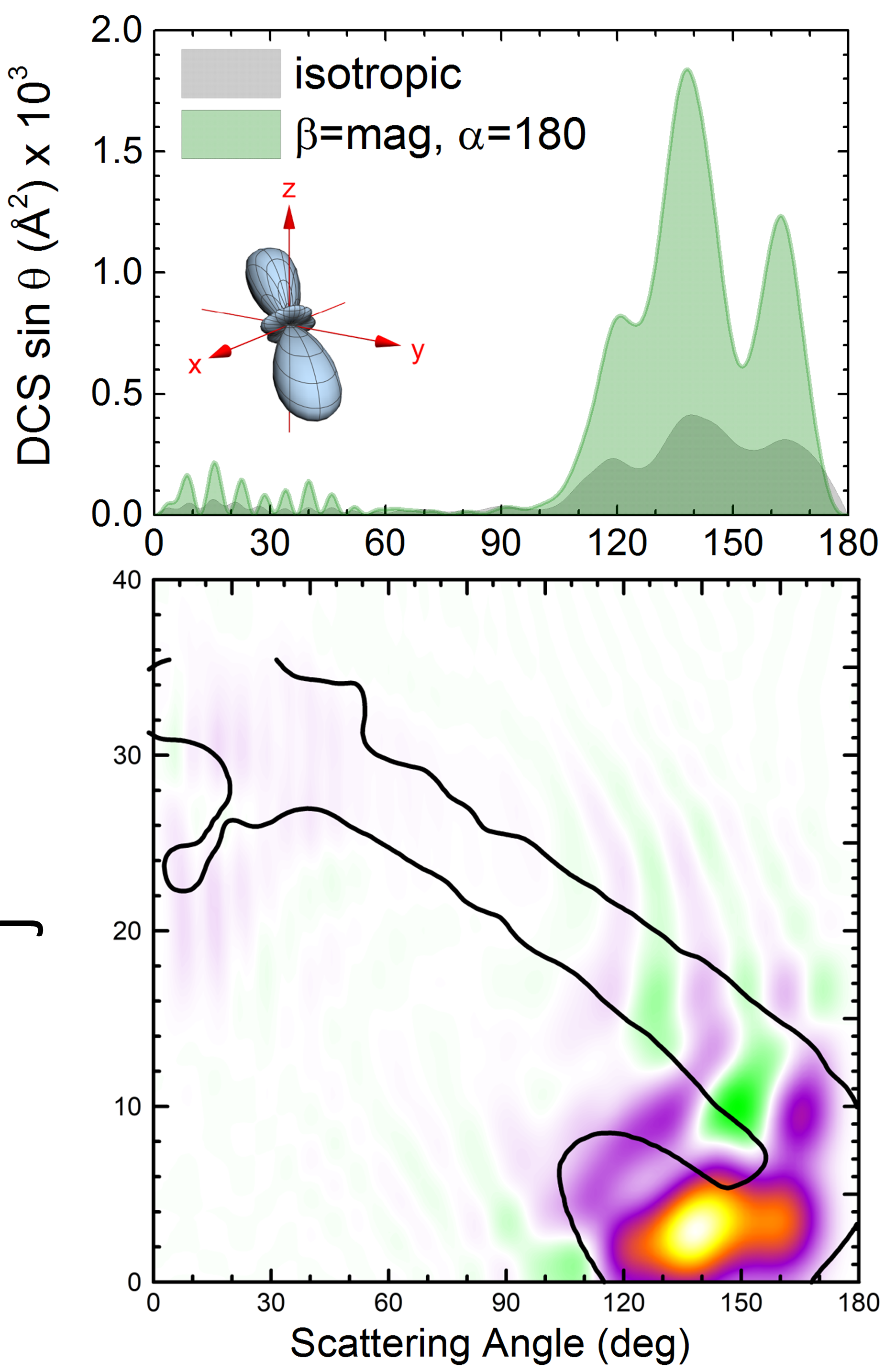}
\caption{As Fig. \ref{Fig2} but for a $\beta={\rm mag}$, $\alpha=180 ^{\circ}$
preparation.} \label{Fig6}
\end{figure}

So far, we have observed how interference between different mechanisms affects the shape
of the DCS for collisions leading to HD($v'$=1,$j'$=0), for initial $j$=0 and $j$=2. The
question that we address in this article is  if it would be possible to modulate the
effect of this interference by using different polarization of the reactants. Using the
analogy with the double-slit experiment, whether we can change not only the intensity,
but also the position of the bands and the nodes of probability. To address this
question, in the remainder of this article we will present and discuss observable DCSs
for different polarizations of the reactants.

\begin{figure*}[t!]
\centering
\includegraphics[width=0.9\linewidth]{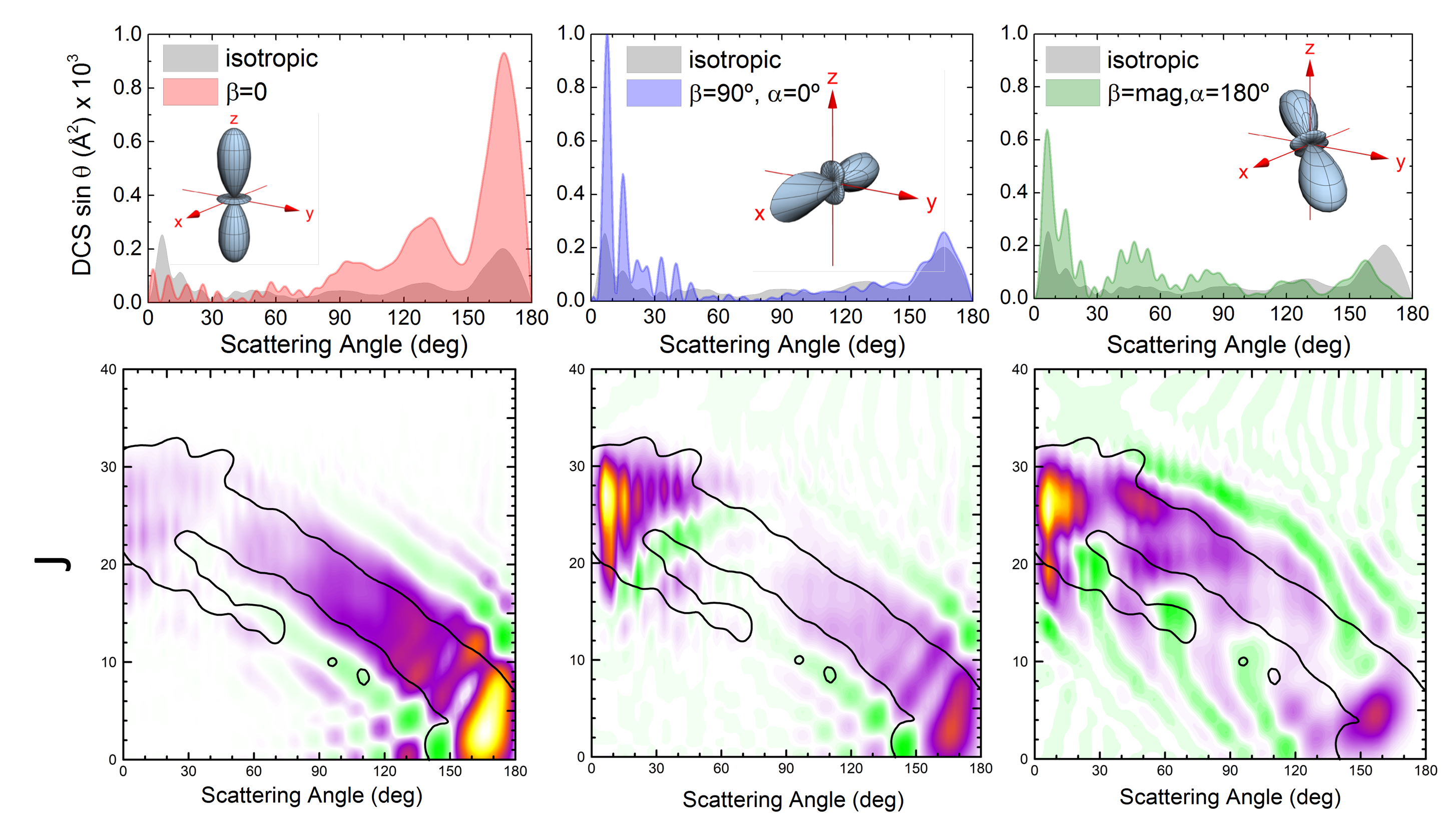}
\caption{Differential Cross Sections (upper panels) for the H +D$_2($v$=0,
$j$=2)$ $\rightarrow$ HD($v'$=3,$j'=0$) + D reaction at $E_{\rm col}=$1.97 eV
for $\beta$=0 (left panel),$\beta=90^{\circ}$, $\alpha=0$ (middle panel), and
$\beta={\rm mag}$, $\alpha=180^{\circ}$ (right panel). The isotropic DCS is
shown in gray and the observable DCSs are shown in red ($\beta=$0), blue
($\beta=90^{\circ}$), and green ($\beta={\rm mag}$) . The extrinsic preparation
of the internuclear axis of D$_2$ referred to the scattering frame are shown in
insets. The corresponding QM GDF for the various polarization schemes are shown
in the bottom panels.}\label{Fig7}
\end{figure*}

Fig.~\ref{Fig2} displays the DCS for $\beta$=0, for which the laboratory frame
$Z$ axis and the initial relative velocity (scattering frame $z$ axis) are
parallel to each other. Under these conditions, D$_2$ internuclear axis is
preferentially along the relative velocity (actually, the DCS for $\beta$=0 is
strictly equivalent to that it would be obtained if all scattering would have
been caused by $\Omega$=0). Therefore, it can can be expected that this axis
preparation would foster the {\em spiral} mechanism, that covers the whole
range of scattering angles, and it is associated with an essentially collinear
transition state. The {${\rm d}\sigma(\theta|\beta=0)$} exhibits three very
sharp peaks with deep minima between them. The position of the peaks are only
slightly shifted with respect to the peaks found in the isotropic DCS. Based on
a QCT analysis,\cite{JAASZ:CS16} the {\em ear} mechanism was assigned to
collisions going through a T-shape transition state \cite{GMW:JCP08,GMWA:JCP08}
and, accordingly, it can be expected that this mechanism will be much less
important for $\beta$=0. To gain further insight onto the changes observed in
the DCS, in the bottom panel of Fig.~\ref{Fig2} the QM GDF is depicted. As
expected, the relative importance of the {\em spiral} mechanism has soared,
whilst the {\em ear} one has almost disappeared. By inspection of the QM GDF,
it can be concluded that the most backward peak is caused by $J$s from 0 to 15,
the second peak comes from partial waves up to 20, and the third peak, at
115$^{\circ}$, is originated by two separate groups of $J$s, one belonging to
the {\em spiral} and the other from the remnant of the {\em ear} mechanism.

\begin{figure*}
\centering
\includegraphics[width=0.8\linewidth]{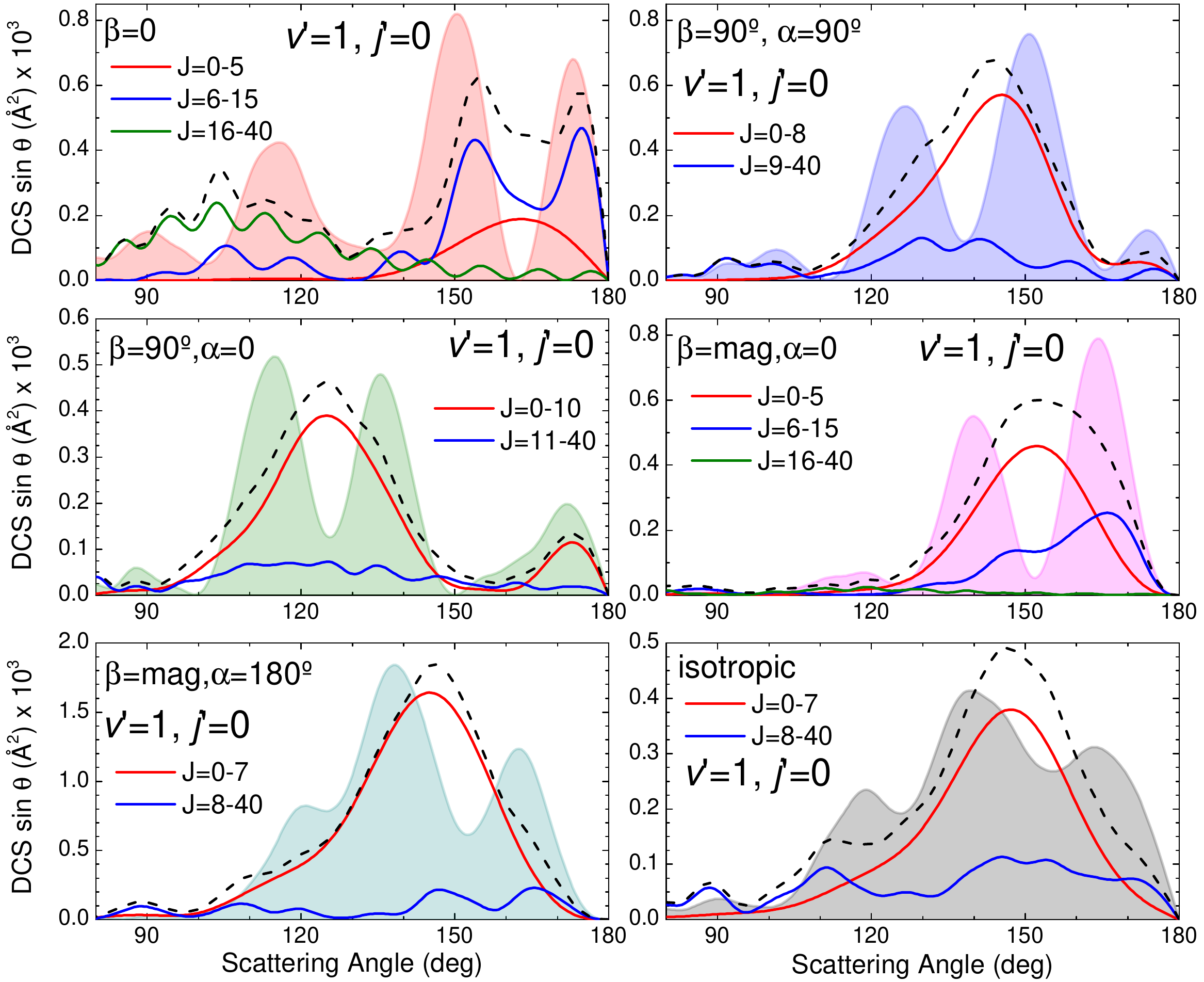}
\caption{Comparison of the different DCS for the H +D$_2$ $\rightarrow$ HD($v'$=1,$j'=0$)
+ D reaction at $E_{\rm col}=$1.97 eV for different experimental set-up. To illustrate
the effect of the interferences in the shape of the DCSs, the contribution of different
groups of partial waves have been calculated (see text for further details). The overall
DCS is shown shaded and the sum of the contributions of the different groups of $J$s is
shown as dashed black lines. }\label{Fig8}
\end{figure*}
The situation significantly changes when we simulate the observable DCS for
$\beta=$90$^{\circ}$, $\alpha=$0, shown in Fig.~\ref{Fig3}. This preparation
corresponds to side-on, coplanar encounters, with the internuclear axis
preferentially along the $x$ direction (on the $\bm k$-$\bm k'$ plane). Again,
we can observed three peaks in the DCS somewhat shifted with respect to the
much smoother peak of the isotropic DCS. In any case, the relative intensity of
those peaks has changed dramatically as compared to isotropic DCS. The peak at
110$^{\circ}$ is considerably larger and isolated from the second peak by a
deep minimum. On the other hand, the most backward peak is considerably less
important, and overall the DCS is considerably less backward. The explanation
of the main features of the DCS can be extracted from the analysis of the
$Q^{90^{\circ}}_{0}(\theta,J)$. As can be observed, in contrast with the
$\beta=$0 case, the {\em ear} mechanism is strongly favoured and becomes more
important than the spiral. The interference between those mechanisms splits the
{\em spiral} in two sharp peaks at $\approx 110^{\circ}$ and $\approx
140^{\circ}$ that are separated by destructive interference (shown as a green
band in the $J-\theta$ map) at 120$^{\circ}$, which is the position of the dip.
Whilst in the isotropic DCS, scattering from the {\em ear} and {\em spiral} are
merged at the most backward region, for DCS($\beta=$90$^{\circ}$, $\alpha=$0)
the {\em ear} is confined into a more sideways region leading to a reduction in
the backward signal as a results of much less interference with the {\em
spiral} mechanism.

Intuitively, side-on collisions should be also fostered with a $\beta=$90$^{\circ}$,
$\alpha=$90$^{\circ}$ axis preparation, and therefore minor changes would be expected
when moving from the $\beta=$90$^{\circ}$, $\alpha=$0 to the $\beta=$90$^{\circ}$,
$\alpha=$90$^{\circ}$ arrangement. However, as it is shown in Fig.~\ref{Fig4}, this is
not the case. For $\alpha=$90$^{\circ}$ we observe three peaks, and whilst the
intensities of the first and the third peak have barely changed, the second peak is
50\,\% larger and, hence, is now the most prominent peak. More interestingly, the two
most sideways peaks are in anti-phase with the peaks observed for $\alpha=$0. As can be
observed in the $Q^{90^{\circ}}_{90^{\circ}}(\theta,J)$, shown in the bottom panel of
Fig.~\ref{Fig4}, those two peaks are caused by interference between the {\em spiral} and
{\em ear} mechanisms, hence proving that the shift in $\alpha$ is enough to change the
phase of the interference causing a conspicuous shift in the position of the peaks of the
DCS. It is also noticeable that the scattering from the {\em ear} mechanism has moved
towards backward scattering and splits in two parts by a destructive interference which
coincides with the deep minimum observed in the DCS.



Next, we will examine the effects produced with preparations with
$\beta=$54.74$^{\circ}$, the magic angle, for which the second Legendre polynomial is
zero and hence the $U^{(2)}_0(\theta)$ PDDCS does not contribute to the observable DCS.
For this configurations, the resulting distribution of the internuclear axis are tilted
with respect to the initial relative velocity. The results for a $\beta=$mag, $\alpha=$0
preparation are shown in Fig.~\ref{Fig5}. Under this preparation the DCS is considerably
more backwards, as the most sideways peak has vanished whilst the intensity of the most
backwards peak is considerably larger. Besides, the minima are deeper. From inspection of
the $Q^{\rm mag}_{0}(\theta,J)$, it is possible to learn that the {\em ear} mechanism
gives rise to products at larger scattering angles and, due to interference between the
two mechanisms, the peak is split in two very sharp peaks at $\theta$=140$^{\circ}$ and
165$^{\circ}$. It is worth noticing that so far we have proved four different
preparations and that the degree of control of the DCS is so subtle that by changing the
preparation, it is possible to select which peak will be the most prominent. For all the
aforementioned cases, the peaks of the DCS are sharper than those from the isotropic DCS.

When a $\beta=$54.74$^{\circ}$, $\alpha=$180$^{\circ}$ preparation is simulated (Fig.
\ref{Fig6}), the shape of the DCS is remarkably similar to the isotropic one. However,
the intensity is much bigger, more than a factor of two, regardless the scattering angle
considered. In fact, whilst the same scale was used to represent the observable DCS in
Figs.~\ref{Fig2}-\ref{Fig5}, we had to double the scale to represent the  observable DCS
for this preparation. Apart form the huge increment of the intensity, inspection of the
$Q^{\rm mag}_{180 \circ}(\theta,J)$ allows us to assign the two peaks to interference
between the {\em ear} and the {\em spiral} mechanisms.

Finally, we will carry out the same analysis for collisions leading to $v'$=3. The most
significant difference between the results of the analysis of the isotropic DCS for the
HD product in $v'$=1 and $v'$=3 is the fact that we have found two competing mechanisms
in the former, the {\em ear} and the {\em spiral}, capable of interfering with each
other, whilst for the $v'$=3 the only surviving mechanism is the {\em spiral},
characterized by a collinear transition state. Coherences will thus be limited to nearby
$J$ values pertaining to the same mechanism. Therefore, the simulation of the DCSs and
their analysis for collisions leading to HD($v'$=3,$j'$=0) furnish us with a
counter-example to that discussed for $v'$=1.

In the absence of significant interference, it can be expected that the shape
of the DCS can be rationalized using simple geometric arguments. Accordingly,
the $\beta=$0 preparation will promote head-on collisions, characterized by
small impact parameters (or small $J$). This is what is observed in the left
panel of Fig.~\ref{Fig7}: the most backward peak increases by almost a factor
of five (the maximum possible enhancement for $j$=2). In contrast, forward
scattering is somewhat diminished, since high impact parameters are less
effective for head-on collisions. Overall, the shape of the DCS, except for the
just mentioned differences, has not changed much and the position of the peaks
are unaltered with respect to the isotropic DCS. In turn, for
$\beta=$90$^{\circ}$, $\alpha=$0$^{\circ}$, side-on collisions with larger
impact parameters would be preferred, fostering forward scattering to values
close to the possible limit (a factor of five for the more forward peak) whilst
backward scattering is almost unaffected. For the $\beta=$54.74$^{\circ}$,
$\alpha=$180$^{\circ}$ preparation, collisions are less side-on and this gives
rise to enhancement in the sideways and forward scattering, although to a less
extent that for $\beta=$90$^{\circ}$. Again, the number and position of the
peaks barely change for the different preparations tested. Inspection of the
respective $Q^{\beta}_{\alpha}(\theta,J)$  confirms this analysis. However, as
an additional information, it is possible to identify the positions of the
minima and maxima in the DCS, and the set of the responsible $J$ values, with
special relevance in forward scattering. Results for $\beta=90^{\circ}$,
$\alpha=90^{\circ}$ and $\beta={\rm mag}$, $\alpha=0$  are not shown because
for HD($v'$=3,$j'$=0) products, the observable DCSs does not change
significantly with $\alpha$.

With the help the  $Q^{\beta}_{\alpha}(\theta,J)$ GDFs, we have
concluded that interference phenomena are very relevant for collisions leading
to HD($v'$=1,$j'$=1) when a polarized distribution of the D$_2$ internuclear
axis is used. To verify that this is certainly the case, in Fig.~\ref{Fig8} we
compare the DCSs calculated for all the partial waves (shadow background) to
the DCS evaluated with restricted groups of partial waves. The boundaries for
the different groups of $J$s were selected to isolate the main structures
observed in the $Q^{\beta}_{\alpha}(\theta,J)$. Finally, a DCS is evaluated as
the (incoherent) sum of the DCSs for each of the different groups (dashed line)
and compared with the actual observable DCSs that includes all the coherences.
Regardless of the experimental preparation, it is evident the big difference
between the sum of the separate contributions (with no interference between
them) and the DCSs that includes all possible coherences. It is thus clear that
interference between the two mechanisms determines the shape of the DCS. It is
worth noticing that neglecting interference between the different groups lead
to a fusion of the main peaks, and that this effect somewhat lessens for the
isotropic internuclear axis distribution. Clearly, the various
preparations of internuclear axis correlate with different combinations of
$\Omega$ values, hence, fostering mechanisms associated with them.
Interestingly, most of the various preparations overcome, at least partially,
the summation over $\Omega$ values that causes a downgrade of the sharp peaks
observed for $j$=0 as $j$ increases.\cite{JAASZ:CS16}

\section{Conclusions}

Based on the present, accurate QM calculations, we have concluded that by an
adequate preparation of polarized ({\em i.e.} anisotropic) distributions of
internuclear axes of the reactants, aligning and orienting the rotational
angular momentum of the reaction before the reaction takes place, it is
possible to change dramatically the shape and intensity  of the observable
differential cross section. Throughout this article, we have simulated the
effect of several prototypical preparations on the differential cross section
for the exchange reaction between H and D$_2$($v$=0,$j=$2) at high collision
energies, where several reaction mechanisms are relevant. Our results show that
changes in the reactants polarization have a dramatic effect on the shape and
intensity of the differential cross section that can be exquisitely controlled.
To ascertain the origin of the different peaks found in the observable DCS, we
have extended the formalism of the generalized quantum deflection function to
calculate joint $\theta$-$J$ quasi-probability distributions for anisotropic
reactant's polarizations, using the polarization moment formalism. Using this
methodology, it has been shown that the resulting interference pattern is very
sensitive to small changes in the polarized distribution of the reactants;
moreover, the interference effects on the differential cross sections are
generally more pronounced than for an isotropic molecular axis distribution. We
believe that these results are rather general as long as several reaction
mechanisms coexist, and it can be expected that the modulation of the
interference pattern with different extrinsic preparations are experimentally
accessible and should be also observed for other reactive or inelastic
processes.

\section{Acknowledgment}

The authors  heartily thank Prof. Enrique Verdasco for his support and help with the
calculations. Funding by the Spanish Ministry of Science and Innovation (grant
MINECO/FEDER-CTQ2015-65033-P) is also acknowledged. P.G.J. acknowledges funding by
Fundacion Salamanca City of Culture and Knowledge (programme for attracting scientific
talent to Salamanca).


\providecommand*{\mcitethebibliography}{\thebibliography}
\csname @ifundefined\endcsname{endmcitethebibliography}
{\let\endmcitethebibliography\endthebibliography}{}

\end{document}